\documentclass{aastex}
\usepackage{emulateapj5}
\usepackage{apjfonts}

\def\h#1{\hbox{${}^{#1}$H}}
\def\he#1{\hbox{${}^{#1}$He}}
\def\li#1{\hbox{${}^{#1}$Li}}
\def\be#1{\hbox{${}^{#1}$Be}}
\def\b#1#2{\hbox{${}^{#1#2}$B}}
\def\c#1#2{\hbox{${}^{#1#2}$C}}

\def\h502{\hbox{$ h^{2}_{50}$}}

\def\la{\mathrel{\mathpalette\fun <}}
\def\ga{\mathrel{\mathpalette\fun >}}
\def\fun#1#2{\lower3.6pt\vbox{\baselineskip0pt\lineskip.9pt
  \ialign{$\mathsurround=0pt#1\hfil##\hfil$\crcr#2\crcr\sim\crcr}}}

\slugcomment{ApJ, Vol. 595, in press}

\shorttitle{Nucleosynthesis in Baryon-Rich GRB Outflows}
\shortauthors{Inoue, Iwamoto, Orito and Terasawa}

\begin{document}

\title{Nucleosynthesis in Baryon-Rich Outflows\\ Associated with Gamma-Ray Bursts}

\author{Susumu Inoue$^{1,2}$, Nobuyuki Iwamoto$^{2,3}$, Manabu Orito$^{2,4}$ and Mariko Terasawa$^{2,5}$}
\affil{$^1$Max-Planck-Institut f\"ur Astrophysik,
       Karl-Schwarzschild-Str. 1, Postfach 1317, 85741 Garching, Germany; inouemu@MPA-Garching.MPG.DE\\
       $^2$Division of Theoretical Astrophysics,
       National Astronomical Observatory,
       2-21-1 Osawa, Mitaka, Tokyo 181-8588, Japan}
\altaffiltext{3}{Present address:
                 Department of Astronomy, School of Science, University of Tokyo,
                 7-3-1 Hongo, Bunkyo-ku, Tokyo 113-0033, Japan; niwamoto@astron.s.u-tokyo.ac.jp}
\altaffiltext{4}{Present address:
                 Research Laboratory for Nuclear Reactors, Tokyo Institute of Technology,
                 2-12-1 Ookayama, Meguro-ku, Tokyo 152-8550, Japan; orito@nr.titech.ac.jp}
\altaffiltext{5}{Present address:
                 Center for Nuclear Study, University of Tokyo,
                 Wako, Saitama 351-0198, Japan; mariko@cns.s.u-tokyo.ac.jp}

\begin{abstract}
Robust generation of gamma-ray bursts (GRBs)
 implies the formation of outflows with very low baryon loads and highly relativistic velocities,
 but more baryon-rich, slower outflows are also likely to occur
 in most GRB central engine scenarios,
 either as ``circum-jet winds'' or ``failed GRBs''.
Here we study the possibility of nucleosynthesis within such baryon-rich outflows
 by conducting detailed reaction network calculations
 in the framework of the basic fireball model.
It is shown that high baryon load fireballs attaining mildly relativistic velocities
 can synthesize appreciable quantities of heavy neutron capture elements
 with masses up to the platinum peak and beyond.
Small but interesting amounts of light elements such as deuterium and boron can also be produced.
Depending on the neutron excess and baryon load,
 the combination of high entropy, rapid initial expansion and gradual expansion at later times
 can cause the reaction flow to reach the fission regime,
 and its path can be intermediate between those of the $r$- and $s$-processes (``$n$-process'').
The nucleosynthetic signature of these outflows may be observable
 in the companion stars of black hole binary systems and in the most metal-poor stars,
 potentially offering an important probe of the inner conditions of the GRB source.
Contribution to the solar abundances for some heavy elements may also be possible.
The prospects for further developments in various directions are discussed.
\end{abstract}

\keywords{nuclear reactions, nucleosynthesis, abundances
      --- gamma rays: bursts
      --- stars: abundances
}

\section{Introduction}
Although many fundamental aspects of the nature of gamma-ray bursts (GRBs) remain mysterious,
 an essential requirement for any type of central engine to successfully generate a GRB
 is the formation of an ultrarelativistic outflow with bulk Lorentz factor $\Gamma \ga$ 100,
 so that the gamma-ray compactness of the emitting region can be sufficiently small.
This in turn implies the realization of a very low baryon load within the outflow,
 typically of isotropic equivalent mass $M \la 10^{-4} {\rm M_\sun}$.
Regardless of its detailed properties,
 the conditions at the base of such an outflow should generically be
 radiation-dominated with very high entropy,
 high temperature ($\ga$ MeV) 
 and very high optical depth 
 (see reviews by Piran 1999, M\'esz\'aros 2002).

The baryonic matter in the outflow is also expected to be enriched in neutrons,
 as the GRB engine is likely to involve
 some kind of collapsed stellar mass object,
 possibly with intense thermal neutrino emission
 (Fuller, Pruet \& Abazajian 2000, Pruet, Woosley \& Hoffman 2003, Beloborodov 2003b).
Matter expanding from MeV-temperature, neutron-rich conditions
 is known to be conducive to fusion of nuclei from initially free protons and neutrons.
It is thus of interest to study the effects and implications of nucleosynthesis inside GRB outflows,
 which was the subject of several recent papers
 (Lemoine 2002, Pruet, Guiles \& Fuller 2002, Beloborodov 2003b).
Production of nuclei turns out to be quite limited in
 the very high entropy ($\ga 10^5 k_B$ per baryon) and extremely rapid expansion ($\la 0.3$~msec)
 characteristic of GRB outflows,
 resulting only in the synthesis of some D and $^4$He.
The low total mass ejected per event
 and the low intrinsic event rate expected for GRBs imply that
 these products themselves are probably difficult to observe.

Besides the baryon-poor, ultrarelativistic outflow triggering the observable GRB emission,
 most GRB progenitors should also give rise to
 associated outflows with much higher baryon-loading and lower velocities under different circumstances.
Mounting evidence points to the high $\Gamma$ GRB outflow
 being narrowly collimated into jets (Rhoads 2001, Frail et al. 2001, Panaitescu \& Kumar 2002).
It is then highly probable that
 an outer, sheath-like wind of slower, baryon-rich material
 surrounds and coexists with the inner, fast jet.
Generically speaking,
 whatever physical mechanism that initiates the GRB jet
 (pair production via neutrino annihilation, magnetohydrodynamic (MHD) stresses, Blandford-Znajek process, etc.)
 must act on regions where the baryon content is tolerably low, presumably nearest to the jet axis
 (e.g. M\'esz\'aros \& Rees 1992, 1997, Levinson \& Eichler 1993).
However, this mechanism is also likely to affect more baryon-contaminated areas
 peripheral to the jet, inducing a baryon-rich, circum-jet flow as a natural byproduct.
For models requiring jet penetration through envelopes of massive stars
 (Woosley 1993, Paczy\'nski 1998, Nakamura 1998, Wheeler et al. 2000, Wheeler, Meier \& Wilson 2002),
 similar flows can additionally occur after jet formation
 through entrainment and mixing with the ambient stellar material
 (Aloy et al. 2000, 2002, MacFadyen, Woosley \& Heger 2001, Zhang, Woosley \& MacFadyen 2003).
Alternatively, if an accretion disk around a black hole is formed,
 as envisioned in many current scenarios
 (Woosley 1993, Popham, Woosley \& Fryer 1999, Fryer, Woosley \& Hartmann 1999, Narayan, Piran \& Kumar 2001),
 the outer parts of the disk may drive a baryon-rich wind
 by mechanisms different from the GRB jet
 (e.g. M\'esz\'aros \& Rees 1997, Narayan et al. 2001, Janka \& Ruffert 2002).
This is anticipated in models invoking rotating core collapse of massive stars
 (e.g. Woosley 1993, MacFadyen \& Woosley 1999, Vietri \& Stella 1998),
 as well as compact binary mergers
 (e.g. Ruffert et al. 1997, Janka et al. 1999, Rosswog \& Ramirez-Ruiz 2002).
Such winds may possibly convey total energies
 comparable to or even larger than the GRB jet itself (Levinson \& Eichler 2000, MacFadyen 2003).
The presence of such circum-jet winds or flows may be consistent with interpretations
 of the statistics of afterglow light curve breaks
 (Rossi, Lazzati \& Rees 2002, Zhang \& M\'esz\'aros 2002).
The recently recognized class of X-ray flashes (Heise et al. 2001, Kippen et al. 2002)
 may also be triggered by such kinds of flows (Zhang et al. 2003).

Baryon-rich outflows may also occur without concomitant GRBs
 in various progenitor scenarios,
 when conditions are such that the baryon-loading process acts more thoroughly,
 or the mechanism for GRB jet production and/or propagation operates less efficiently
 (e.g. Woosley 1993, Narayan et al. 2001, Janka \& Ruffert 2002,
 Wheeler et al. 2000, 2002, Woosley, Zhang \& Heger 2002b, MacFadyen 2003).
Such ``failed GRBs'' may have event rates higher than successful GRBs
 (Dermer 2000, Woosley et al. 2002b, Huang, Dai \& Lu 2002).
Depending on the identity of the actual  GRB engine,
 perhaps some identification of failed GRBs can be made with observed ``hypernovae''
 (i.e. explosions similar to ordinary supernovae but with higher energies;
 Iwamoto et al. 1998, Nomoto et al. 2002, Wheeler et al. 2002, Woosley et al. 2002b),
 X-ray flashes (e.g. Dermer, Chiang \& Mitman 2000),
 or with some hitherto undiscovered type of transient
 (e.g. Li \& Paczy\'nski 1998, MacFadyen et al. 2001). 
They may also comprise transitory phenomena
 preceding or ensuing a phase of genuine GRB emission,
 possibly appearing as X-ray precursors (Murakami et al. 1991) or X-ray tails (Yoshida et al. 1989)
 observed in some GRBs.
Either as circum-jet winds or failed GRBs,
 these baryon-rich outflows (hereafter BROs) should be typified by
 lower entropy, slower expansion and higher ejecta mass compared to GRB jets,
 which are all more propitious from a nucleosynthesis viewpoint.
\footnote{
Such phenomena are sometimes called ``dirty fireballs'' in the GRB literature,
 but this term is avoided here,
 as these outflows can be aesthetically appealing for nucleosynthesis.}

Besides the production of light elements,
 a particularly interesting issue is whether appreciable heavy element synthesis
 by neutron capture reactions
 can take place in such environments.
Elements heavier than Fe in the solar system
 are believed to arise from two distinct types of neutron capture processes,
 the $r$-process, for which the timescale of neutron capture is much faster than $\beta$-decay,
 and the $s$-process, for which the relation is vice-versa
 (see reviews by Cowan, Thielemann \& Truran 1991, Meyer 1994, Qian 2003).
The $s$-process proceeds with a nuclear reaction flow close to the line of $\beta$-stability,
 and is favored in situations with relatively low neutron abundances.
Its main astrophysical site is thought to be asymptotic giant branch stars,
 a very different environment compared to GRB-BROs.
In contrast, exposure to high neutron fluxes is necessary for the $r$-process, 
 in which the reaction flow occurs far into the neutron-rich region close to the neutron dripline.
Currently, the most widely discussed $r$-process site is the neutrino-driven winds of core-collapse supernovae
 (e.g. Woosley et al. 1994, Takahashi, Witti \& Janka 1994, Terasawa et al. 2002, Terasawa 2002,
 Woosley, Heger \& Weaver 2002a).
Efficient operation of the $r$-process in such winds
 requires high entropy ($\ga 100-200k_B$ per baryon), rapid expansion ($\sim 1-100$~msec)
 and high neutron excess ($Y_e\la$ 0.43 in terms of the electron fraction).
GRB-BROs are intriguing in this regard, as their physical properties should be similar in a number of ways.
The typical conditions in relativistic BROs can be even more extreme,
 but depending on the baryon load and other parameters,
 a successful $r$-process may also be possible.
While this has been the subject of speculation by various authors in different contexts
 (Levinson \& Eichler 1993, Ruffert et al. 1997, MacFadyen et al. 2001,
 Lemoine 2002, Pruet et al. 2003),
 quantitative nucleosynthesis calculations in the relevant physical conditions have not been performed.
(See \S 5 for a discussion relating to previous studies of the $r$-process
 in jet-driven supernova explosions.)
Pruet et al. (2003; see also Beloborodov 2003b) recently evaluated the degree of neutronization expected
 in GRB central engines based on black hole accretion disk models,
 and discussed the implications for nucleosynthesis in winds emanating from such disks,
 but no reaction network calculations were carried out.

This work is an exploratory investigation of nucleosynthesis in BROs associated with GRBs
 and their observational consequences.
We utilize the simplified dynamical framework of the basic fireball model,
 accounting for the crucial coasting phase (\S 2.1).
A key feature is the incorporation of large and detailed nuclear reaction networks 
 including light, neutron-rich nuclei (\S 2.2). 
The treatment in this paper is limited to the relativistic regime,
 which may apply to BROs in compact binary mergers or supranovae
 (but probably not to those occurring inside massive stars).
Although this is a first step in a potentially broader problem,
 we find interesting results with regard to both light and heavy element production (\S 3).
Some important observational implications of these nucleosynthetic products are pointed out,
 particularly for understanding the nature of the ill-understood GRB central engine (\S 4).
We conclude with a discussion of various interesting possibilities for future studies (\S 5).

\section{Model Description}

\subsection{Dynamics}

We start with the simplest possible dynamical model:
 a spherical, adiabatic, thermally-driven flow that is
 freely expanding and in steady state (Paczy\'nski 1986, 1990).
The key parameters in this wind fireball model are
 the luminosity $L = 10^{52} L_{52} {\rm erg \ s^{-1}}$,
 the initial radius $r_0 = 10^7 r_{0,7} {\rm cm}$,
 and the dimensionless entropy $\eta = L/\dot M c^2$,
 where $\dot M \simeq 5.6 \times 10^{-3} {\rm M_\sun s^{-1}} L_{52} \eta^{-1}$
 is the baryon-loading rate.
The initial bulk velocity is taken to be subrelativistic, $\Gamma_0 \simeq 1$.
Assuming that the available energy is entirely dissipated into thermal form
 at the base of the outflow,
 the fireball plasma is radiation-dominated if $\eta \ga 1$,
 and its initial temperature is
 $k_B T_0 = \left(L/4 \pi r_0^2 c a g\right)^{1/4}
          \simeq 0.93 {\rm MeV} \ L_{52}^{1/4} r_{0,7}^{-1/2}$
 where $a$ is the radiation constant
 and $g=11/2$ the degrees of freedom including photons and electron-positron pairs.
Its optical depth is very high due to the presence of the pairs.
The entropy per baryon $s$ is conserved in the flow and related to $\eta$ and $T_0$ as
 $s/k_B = 4 \eta m_N c^2 /3 k T_0 \simeq 1250 \eta (T_0/{\rm 1 MeV})^{-1}$,
 where $m_N$ is the nucleon mass (Fuller et al. 2000).
The fireball starts to expand adiabatically
 and first undergoes an acceleration phase,
 where the bulk Lorentz factor $\Gamma$ grows with radius
 as $\Gamma \propto r$
 while the comoving temperature $T$ falls as $T \propto r^{-1}$.
This continues until the internal energy is fully converted
 to the bulk kinetic energy of the loaded baryons at the saturation radius $r_s = r_0 \eta$.
The fireball then enters a coasting phase with constant velocity $\Gamma = \eta$,
 where $T \propto r^{-2/3}$.
These simple scalings derive from conservation of energy and rest mass
 along with adiabaticity and are strictly valid for $\Gamma \gg 1$.
Lorentz transformations to the comoving frame give the 
 profiles (trajectories) for the temperature $T$ and baryon density $\rho_b$
 with respect to time $t$:
 for the acceleration phase ($t \le t_s$),
 \begin{eqnarray}
 T(t) &=& T_0 \exp (-t/t_{d0}) \nonumber \\
 \rho_b(t) &=& \rho_{b,0} \exp (-3t/t_{d0}) ;
 \label{eqn:trajacc}
 \end{eqnarray}
 for the coasting phase ($t \ge t_s$),
 \begin{eqnarray}
 T(t) &=& {T_0 \over \eta} \left(1+{t-t_s \over t_{d0}}\right)^{-2/3} \nonumber \\
 \rho_b(t) &=& {\rho_{b,0} \over \eta^3} \left(1+{t-t_s \over t_{d0}}\right)^{-2} .
 \label{eqn:trajcoa}
 \end{eqnarray}
Here $t_{d0} = r_0/c \simeq 0.33 {\rm ~msec} \ r_{0,7}$
 is the initial light crossing time or dynamical time,
 $t_s = t_{d0} \ln \eta$ is the comoving saturation time,
 and $\rho_{b,0} = L/4 \pi c^3 r_0^2 \eta
 \simeq 3.0 \times 10^5 {\rm g~cm^{-3}} \eta^{-1} L_{52} r_{0,7}^{-2}$
 is the initial baryon density.
Note that $T = T_0 \eta^{-1} (t/t_{d0})^{-2/3}$
 and $\rho_b = \rho_{b,0} \eta^{-3} (t/t_{d0})^{-2}$ asymptotically for $t \gg t_s$.
The trajectories above are in fact slightly modified
 by the effect of pair annihilation at $T \la$ MeV, which is included in our calculations.

We see that the acceleration phase is characterized by an exponential expansion
 with e-folding time equal to the dynamical time at the outflow base (Fuller et al. 2000),
 whereas the coasting phase corresponds to a power-law expansion.
A high $\eta$ (low baryon load) fireball
 suffers from a long period of exponential drop in $T$ and $\rho$
 in addition to a very high entropy, both severely hindering nucleosynthesis.
The ensuing coasting phase is then irrelevant 
 and can be neglected for nucleosynthesis in the high $\eta$ outflows of successful GRBs
 (Lemoine 2002, Pruet et al. 2002, Beloborodov 2003b).
In contrast, a low $\eta$ (high baryon load)
 would allow the fireball to reach the power-law regime in a short time,
 where the more gradual decrease in $T$ and $\rho$ along with a lower entropy
 can greatly facilitate nucleosynthesis.
We note that the forms of these trajectories
 are not specific to GRB-related flows,
 but are generic descriptions for adiabatic, radiation-dominated expansion reaching a terminal velocity.
Eqs.\ref{eqn:trajacc} and \ref{eqn:trajcoa}
 can also approximate the trajectories for neutrino-driven winds in core collapse supernovae
 (e.g. Qian \& Woosley 1986, Thompson, Burrows \& Meyer 2001).
The main difference here is in the range of values explored for the physical parameters,
 such as the entropy, density and expansion timescale (see also Meyer 2002).

The description above is also applicable for an outflow confined to a conical channel
 so long as the opening half-angle $\theta \ga \Gamma^{-1}$,
 in which case $L$ and $\dot M$ should be interpreted as isotropic equivalent quantities.
Flows which are strongly collimated, e.g. by external pressure gradients,
 would possess somewhat different scalings of physical variables with radius
 (Lemoine 2002, Beloborodov 2003b).
Note that when compared to the narrow GRB jets,
 the BROs may inherently have a much wider geometry (Levinson \& Eichler 2000),
 so that the present formulation may be useful even for modest Lorentz factor flows.

As is apparent from the above arguments and also verified by actual calculations below,
 the lower the value of $\eta$, the more advantageous it is for nucleosynthesis.
This is in terms of both the reaction flow to heavier nuclei and the total product mass.
However, the above dynamical description
 must be modified if $\eta \la 1$,
 i.e. the baryon load is sufficiently large that the fireball becomes nonrelativistic.
Here we will limit ourselves to the relativistic scalings
 while imposing a lower limit of $\eta = 2$ (i.e. a mildly relativistic fireball),
 which corresponds to total baryon mass $M_b = 2.8 \times 10^{-2} {\rm M_\sun} E_{53}$
 for total outflow energy $E=10^{53} E_{53} {\rm erg}$, in isotropic equivalent terms.
This is on two accounts.
First, the physical regime of completely non-relativistic flows
 would be rather distinct from the high $\eta$ GRB fireballs,
 and their actual properties, if they occur, are more uncertain.
Second, for relatively baryon-free progenitor models
 such as neutron star mergers (Ruffert et al. 1997, Janka \& Ruffert 2002)
 or possibly supranovae (Vietri \& Stella 1998),
 $M_b \simeq 10^{-2} {\rm M_\sun}$ may be a reasonable upper bound
 to the potential amount of baryon contamination (e.g. Woosley 1993).
As the assumption of free expansion should also suit these scenarios,
 the simple formulation employed here may be an adequate approximation
 for BROs in such models.
(However, BROs of the ``disk wind'' type may have $T_0$, $s$ and $t_{d0}$
 different from the simple fireball prescription; \S 5).
Cases of much higher baryon loading can happen in other models like collapsars,
 where the outflows form and propagate deep inside massive stars,
 and $M_b \simeq 1 {\rm M_\sun}$ or more may be realized
 (MacFadyen \& Woosley 1999, MacFadyen 2003, Wheeler et al. 2000, 2002),
 but this will be considered elsewhere.
Other effects that can be important for realistic GRB-BRO dynamics
 are touched upon in \S 5.

\subsection{Microphysics}

A critical quantity influencing the nuclear composition of baryonic matter
 is the neutron excess, commonly described in terms of the electron fraction
 $Y_e = (n_{e^-}-n_{e^+})/n_b$, i.e. the net number of electrons per baryon.
The value of $Y_e$ at the base of the outflow should be predictable
 if reliable physical models of the central engine can be constructed, 
 but this is currently a difficult task.
The material in the collapsing cores of massive stars
 typically have $Y_e \la 0.5$,
 whereas material near the surfaces of cold neutron stars have $Y_e \simeq 0.1$.
However, these can be significantly altered by capture of electrons (positrons)
 as well as electron-type neutrinos (antineutrinos) on protons (neutrons)
 during the course of the central engine evolution.
Pruet et al. (2003) and Beloborodov (2003b)
 have recently evaluated the $Y_e$ values that may be realized
 in black hole accretion disk models appropriate for GRB central engines
 (Popham et al. 1999).
Pruet et al. give numbers ranging from $Y_e \la 0.1$ to $Y_e \ga 0.5$
 at the innermost edge of the disk,
 depending on parameters such as the viscosity, accretion rate,
 mass and angular momentum of the central black hole.
\footnote{
These may be further modified by inclusion of neutrino-trapping effects,
 which become crucial at high densities (Di Matteo, Perna \& Narayan 2002).}
Models invoking fireball formation in the envelopes of supermassive stars
 can instead be highly proton-rich, $Y_e \la 1$
 (Fuller \& Shi 1998).
In view of these uncertainties,
 we will treat $Y_e$ as an additional free parameter in this work.
Except when $T_0 \gg$ MeV,
 electron/positron captures within the fireball
 will not be fast enough to change $Y_e$ from its initial value
 (Fuller et al. 2000, Beloborodov 2003b).

For $T_{0}$ and $\rho_{b,0}$ typical of GRB outflows,
 the baryons should initially be in nuclear statistical equilibrium (NSE)
 consisting predominantly of free protons and neutrons,
 so the assembly of nuclei always begins from the lightest elements.
A key feature of the nuclear reaction network codes
 implemented here is the inclusion of a large number of reactions 
 involving light, very neutron-rich nuclei
 (Terasawa et al. 2001, Orito \& Iwamoto 2003, in preparation).
Besides the well-known paths of
 $\alpha$($\alpha \alpha$,$\gamma$)$^{12}$C and
 $\alpha$($\alpha$n,$\gamma$)$^9$Be($\alpha$,n)$^{12}$C
 bridging the mass gaps at $A=5$ and $A=8$,
 new reaction channels allowed via these light neutron-rich nuclides
 have been shown by Terasawa et al. (2001)
 to critically influence the synthesis of even the heaviest nuclei,
 especially for expansion with short timescales and/or high neutron excesses.
This is indeed the case for our conditions here.
Most previous network codes including only a limited number of such reactions
 would give erroneous results for the present problem.
In total, our code used for heavy element synthesis calculations
 includes over 3000 nuclides, mainly on the neutron-rich side of stability.

On the other hand,
 we only deal with a limited number of heavy nuclei on the proton-rich side,
 which, if fully included,
 may possibly affect the reaction flow when $Y_e$ is near 0.5.
We also choose to neglect fission,
 the detailed physics of which is currently very uncertain (e.g. Goriely \& Arnould 2001, Cameron 2001),
 but which may in reality have important consequences
 for some parameter regimes studied here.
These microphysical aspects will be accounted for in subsequent studies (\S 5).
We do include the effects of the increase in temperature and entropy
 from electron-positron pair annihilation occuring at $T \la$ MeV (e.g. Meyer \& Brown 1997).

In many GRB models, strong neutrino irradiation from the central source
 can also affect the nuclear composition (e.g. Beloborodov 2003b),
 but this will not be treated here as it is highly model-dependent (see \S 5).
We remark that
 for all cases under consideration here
 the neutrons remain kinematically well-coupled to the rest of the fireball plasma
 throughout the duration of nucleosynthesis 
 (c.f. Derishev, Kocharovsky \& Kocharovsky 1999, Fuller et al. 2000).

\section{Results and Discussion}

The main parameters in the present problem are $L$, $r_0$, $\eta$ and $Y_e$.
GRB observations put reasonable constraints on $L$ and $r_0$,
 so we will keep these to our fiducial numbers of 
 10$^{52}$ erg~s$^{-1}$ and 10$^7$ cm respectively
 (hence $kT_0 \simeq 0.93 {\rm MeV}$;
 see however \S 5).
We thus concentrate on the effect of variations in $\eta$ and $Y_e$,
 for which a wide range of values can be realized.
The trajectories are very different
 for the limiting cases of high and low $\eta$.
For $\eta = 100$ typical of successful GRBs, $s/k_B \sim 10^5$,
 and the acceleration phase lasts for $t_s \simeq 1.5$~msec
 during which $T$ and $\rho_b$ fall extremely rapidly by 2 and 6 orders of magnitude, respectively.
In comparison, for $\eta=2$, $s/k \sim 2500$,
 and the coasting phase is entered at $t_s \simeq 0.23$~msec
 when $T$ and $\rho_b$ are lower by just factors of 2 and 8, respectively.
Much of the reactions can then proceed during the power-law expansion, especially neutron-captures.

Since the production of the more common $\alpha$ and Fe-group elements at intermediate $A$
 should be dominated by normal stars and supernovae in the Universe,
 the signatures of GRB-BRO nucleosynthesis are to be sought
 in other elements which are not readily synthesized in ordinary stellar environments
 (see however MacFadyen 2003).
We therefore focus our interest here
 on the light element region of D, Li, Be and B,
 and on the neutron capture element region much heavier than Fe.

\subsection{Light Elements}

For calculating light element production, a relatively small network
 limited to nuclei with mass number $A \le 18$ suffices,
 as the reaction flow to higher $A$ is not very large. 
Here we utilize a code developed for big bang nucleosynthesis (BBN),
 which includes the relevant weak interaction processes
 and has been extended and updated with reactions for 40 light nuclei in total
 to study inhomogeneous BBN (Orito \& Iwamoto 2003, in preparation).
The relative ease of these computations allows us to survey 
 large regions of parameter space.

We first discuss the yields of \he4 and D for a wide range of $\eta$.
These elements can be produced appreciably
 even in the high $\eta$ fireballs of successful GRBs
 (Lemoine 2002, Pruet et al. 2002, Beloborodov 2003b),
 and we extend the regime of consideration down to $\eta=2$.
Like BBN, \he4 is the dominant final species besides remaining free nucleons
 for all cases here, as the majority of the available protons (neutrons)
 can become locked up in $\alpha$ particles for neutron-rich (proton-rich) conditions.

\begin{figure}
\epsscale{0.47}\plotone{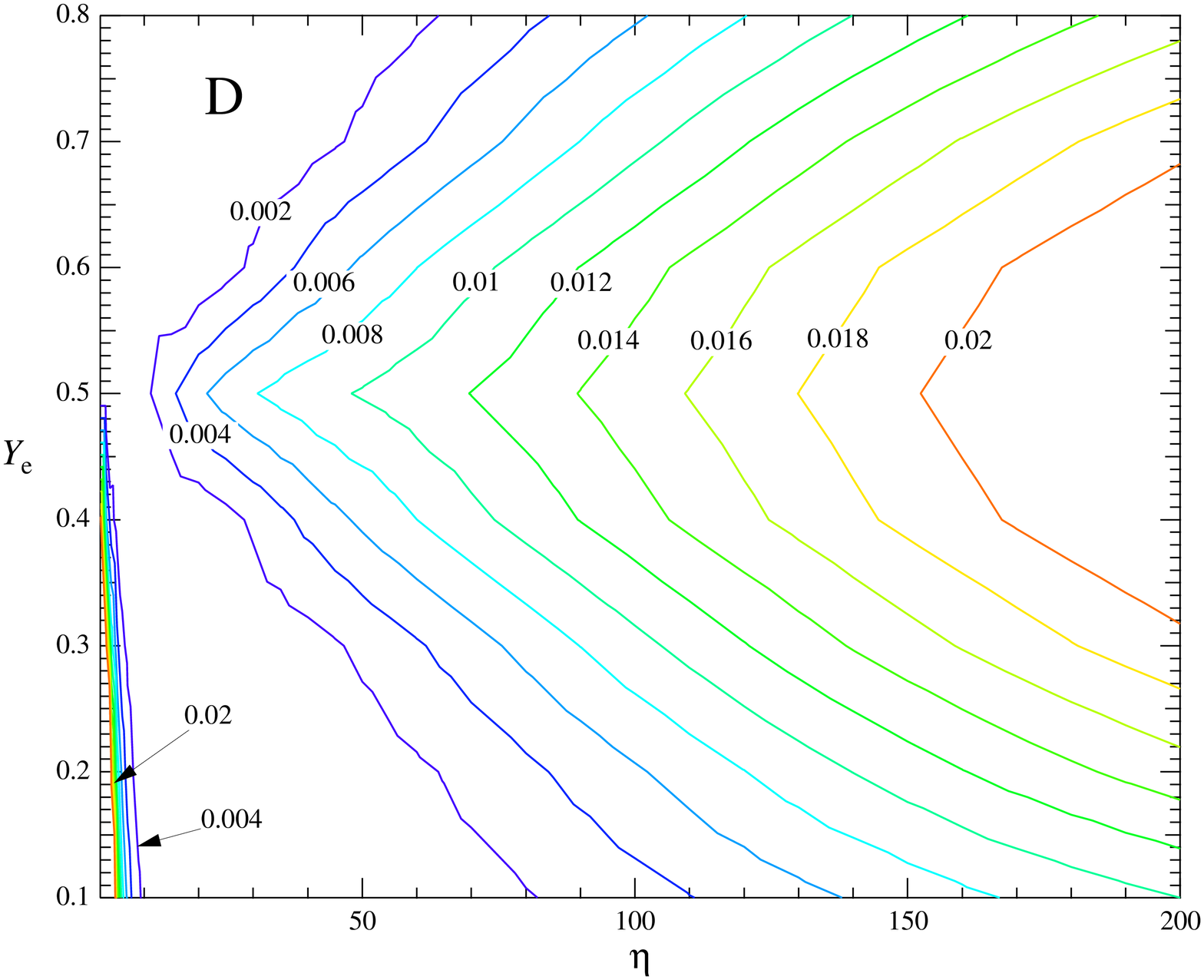} 
\epsscale{0.43}\plotone{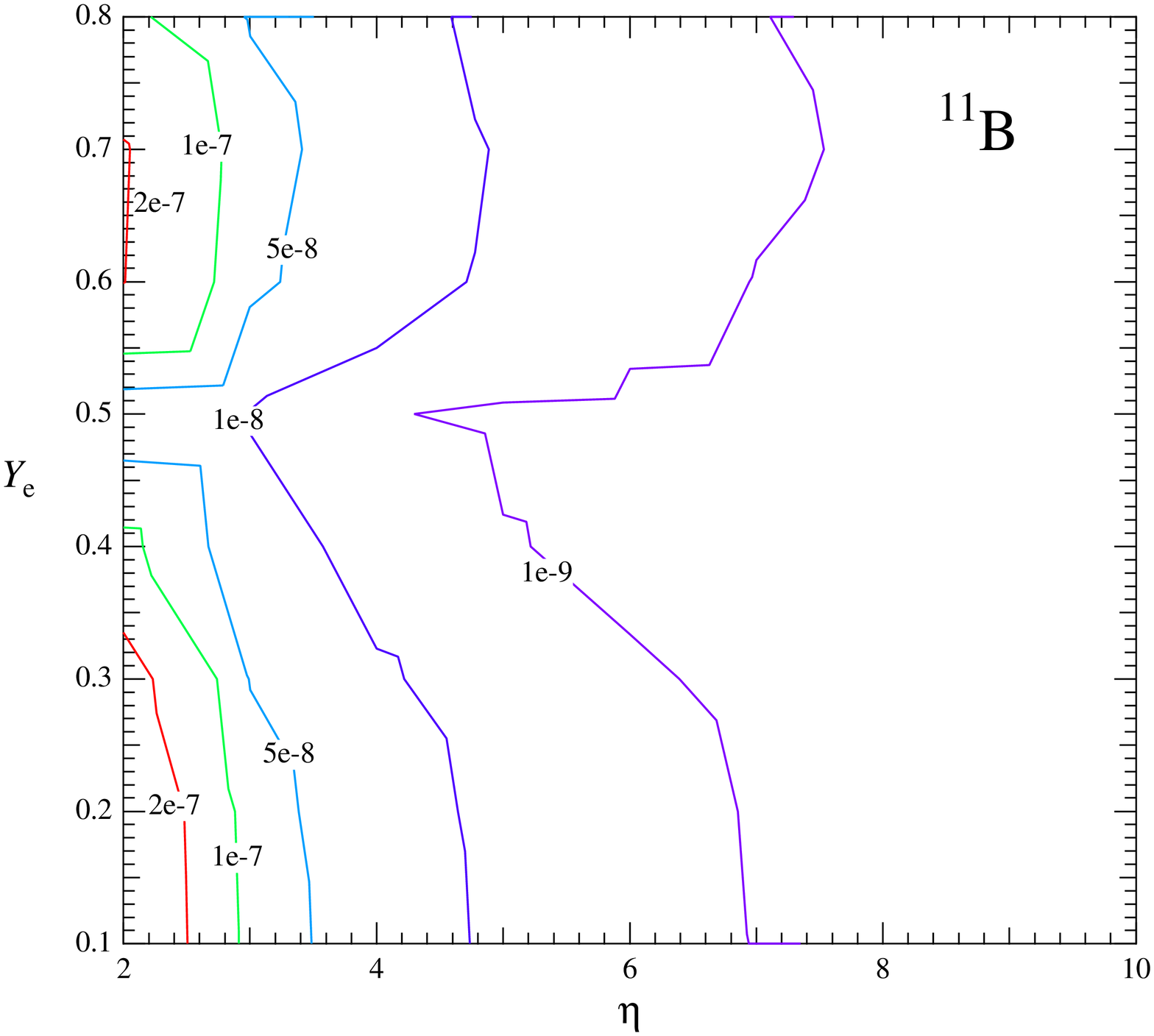} 
\caption{
Contours of the final abundances by mass fraction
 of deuterium (left panel) and boron (right panel)
 for different parameter values of $\eta$ and $Y_e$.}
\label{fig:light}
\end{figure}

The left panel of Fig.~\ref{fig:light} shows the final abundance of D
 as contours of constant mass fraction in the $\eta$-$Y_e$ parameter plane.
For $\eta \ga 20$, the reactions mainly occur during the exponentially expanding acceleration phase. 
D production depends on $\rho_{b,0} \propto \eta^{-1}$
 such that its final abundance is smaller for lower $\eta$ due to larger reaction flows to $\he4$,
 similar to the case of BBN.
The production rate is also proportional to $Y_pY_n \approx Y_e(1-Y_e)$,
 with a maximum at $Y_e = 0.5$. 

Contrastingly,
 for $\eta \la 10$,
 the nucleosynthesis proceeds mainly during the coasting phase with power-law expansion,
 where neutron capture reactions can be active until very late times, $t \la 10^3$~s
 (see \S 3.2).
Here D synthesis can occur via a novel process:
 when $Y_e \la 0.5$, free neutrons remain abundant in the outflow
 until they start decaying into protons,
 which subsequently can undergo the p(n,$\gamma$)d reaction
 during the latest stages of the expansion.
In this regime, D production does not depend strongly on $Y_e$,
 as long as $Y_e \la 0.5$.
This is an interesting D production mechanism that has not been discussed in previous contexts.

Notable amounts of nuclei with $A>$ 4 such as Li, Be and B are created only when $\eta \la 10$.
We display the results for \b11 in the right panel of Fig.~\ref{fig:light}
 for the range of $\eta=2-10$.
Under neutron-rich conditions ($Y_e \le 0.5$),
 \b11 is produced by the reaction chains
 \begin{eqnarray*}
 t(\alpha,\gamma)\li7(n,\gamma)&\hspace{-0.6cm}\li8(n,\gamma)&\hspace{-0.6cm}\li9(\beta)\be9(n,\gamma)^{10}{\rm Be}(n,\gamma)\b11\\
 &\hspace{-0.1cm}\li8(\alpha,n)\b11.&
 \end{eqnarray*}
In proton-rich environments ($Y_e \ge 0.5$),
 the main path for $\b11$ production is
 $\he3(\alpha,\gamma)\be7(\alpha,\gamma)\c11(\beta)$.
Other light elements such as \li6, \li7 and \be9 can also be generated
 at levels of $X \sim 10^{-8}-10^{-7}$.

We remark that under certain conditions,
 light element production may be considerably enhanced through nonthermal reactions
 such as spallation and/or photodisintegration,
 but these effects are outside of our present scope (see \S 5).

\subsection{Heavy Elements}

The synthesis of heavy elements through neutron capture is the focal point of this work.
Such processes obviously encompass a vastly larger number of nuclides
 compared to light element production,
 so full network calculations are much more cumbersome.
Parameter studies with the small, light element network discussed above
 indicate that the flow to heavy nuclei ($A>$16) becomes considerable only for $\eta \la 4$.
We thus concentrate here only on the most favorable case of $\eta=2$,
 and choose selected values for $Y_e$ ($<0.5$).
Using the full network described in Terasawa et al. (2001),
 the calculations below have been followed until all reactions have frozen out,
 typically occurring at $t \sim 10^3$~s.

The $\eta=2$ outflow is characterized by very high entropy ($s/k_{B} \simeq 2500$),
 low density ($\rho_{b,0} \simeq 1.5 \times 10^5 {\rm g~cm^{-3}}$),
 and very short initial expansion timescale ($t_{d0} \simeq 0.33$~ms),
 which can be compared with the values typical of SN neutrino-driven winds (\S 1).
These conditions lead to heavy element synthesis
 which is markedly different from the usual $r$-process in a number of respects.
Displayed in Figure~\ref{fig:heavy} are
 the final abundances by number as a function of mass number,
 for the cases of $Y_e=$ 0.1, 0.3, 0.4, 0.48 and 0.498.
The abundance patterns are clearly discrepant from the solar distribution,
 shown here with arbitrary normalization as the uppermost curve.
When $Y_e \la$ 0.4, implying copious supplies of neutrons,
 considerable reaction flows can occur to high mass numbers,
 ending up in large fractions of the heavier nuclei.
However, we also see that the abundances are at levels of $Y \simeq 10^{-6}$ even at the peaks,
 which is quite low compared to the SN wind $r$-process.
This is due to the combination of small $t_{d0}$,
 which limits the initial, charged particle reaction flow at small $A$,
 and the low density, which hinders the subsequent neutron capture flow to higher $A$.
Note that the apparent abundance peak at $A \simeq$ 250
 is an artifact of our neglect of fission
 and results from artificial pileup at the limit of our reaction network.
In situations where this peak becomes dominant,
 the realistic effects of fission (and possible consequent fission cycling)
 may significantly affect the final abundance pattern (\S 5).
For $Y_e=$ 0.48 and 0.498,
 the flows to high mass nuclei are drastically reduced due to the low neutron excess,
 and only small amounts of elements result at $A \la 50$ and $A \la 20$, respectively.

\begin{figure}
\epsscale{0.9}
\plotone{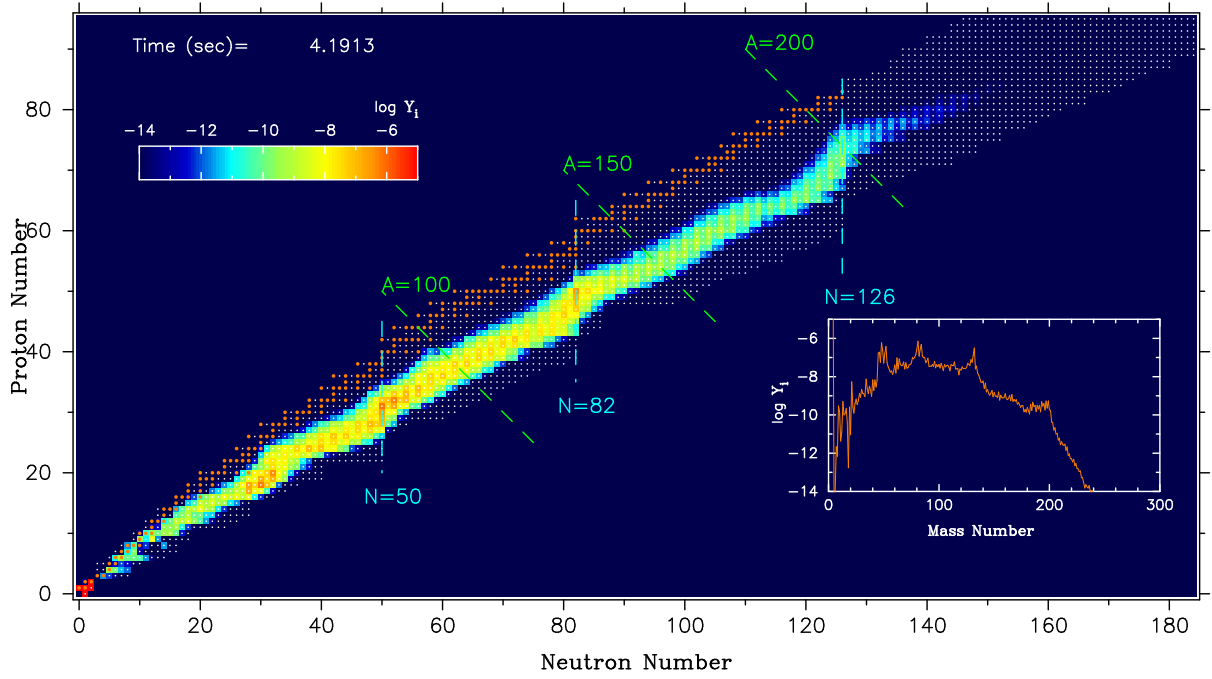} 
\caption{
Snapshot of the reaction flow in the plane of neutron number $N$ versus proton number $Z$
 for the case of $Y_e=0.3$, at time $t \simeq$ 4.2~s and density $\rho_b \simeq 1.2\times 10^{-4}$~g cm$^{-3}$. 
The large and small circles on the grid indicate stable and unstable nuclei, respectively.
The inset shows the number fraction $Y$ of heavy elements as a function of mass number $A$.}
\label{fig:path}
\end{figure}

Our abundance distributions for $Y_e \la$ 0.4 show three peaks around $A \simeq$87, 135 and 205
 located in between the solar $s$- and $r$-process peaks.
This is a clear manifestation of nuclei with neutron magic numbers
 forming on a reaction flow path intermediate between those of the $s$- and $r$-processes.
Figure~\ref{fig:path} shows a snapshot of the calculated reaction flow in the $N$-$Z$ plane 
 for the case of $Y_e=$ 0.3,
 at $t \simeq$ 4.2~s when the density has fallen to 1.2 $\times 10^{-4}$~g cm$^{-3}$. 
Unlike the $r$-process in SN winds,
 the neutron capture rates here are too low to effectively establish
 (n, $\gamma$)-($\gamma$, n) equilibrium due to the low density,
 even when the neutrons constitute a major fraction of the baryons.
Instead, as $\rho_b$ and $T$ decrease steadily in the coasting phase,
 $\beta$-decays begin to compete with the neutron capture reactions,
 leading to a flow path that gradually evolves in time
 from near the dripline ($r$-process-like) to that near the
 $\beta$-stability line ($s$-process-like).

We also note that the differences in the abundances between the peak and interpeak elements
 are larger than that of the solar pattern.
This is due to the fact that nuclei in the moderately neutron-rich region
 have longer $\beta$-decay lifetimes than compared to those near the dripline.
Therefore, reaction flows at the magic nuclei are halted for a longer time,
 resulting in larger peak-to-interpeak contrasts.

A final point is that
 the low density here keeps the neutrons from ever being exhausted by neutron capture,
 in contrast to the $r$-process in SNe.
Instead the neutrons remain abundant after the capture reactions freeze out
 until they freely decay into protons.

Neutron capture nucleosynthesis in environments with low neutron abundances,
 where the flow path is determined by the balance of neutron captures and $\beta$-decays
 rather than (n,$\gamma$)-($\gamma$,n) equilibrium,
 has been discussed previously by Blake \& Schramm (1976) and Blake et al. (1981)
 and christened the ``$n$-process''.
Our physical conditions are quite different
 from those considered by the above authors, in that
 we have a very large neutron abundance but also a low total density.
However, the basic physical processes are similar. 
BROs associated with GRBs
 can therefore be natural astrophysical sites for operation of the $n$-process.

\section{Observational Implications}

The calculations presented above are based on a number of simplifications,
 so they should be regarded as exemplary and tentative expectations
 from nucleosynthesis in actual GRB-BROs.
In particular, they may be fair representations for at least some types of BROs
 in neutron star mergers or supranovae,
 but probably not for BROs triggered inside massive stars (\S 2). 
Various realistic physical effects can modify these results in important ways (\S 5).
Nevertheless, some qualitative trends seen here should remain generally valid.
Below we discuss the observational implications of GRB-BRO nucleosynthesis,
 taking at face value the results of our $\eta=2$, mildly relativistic fireball as a fiducial case.
These considerations should also serve as a reference point
 for more detailed investigations in the future.

\begin{figure}
\epsscale{0.85}
\plotone{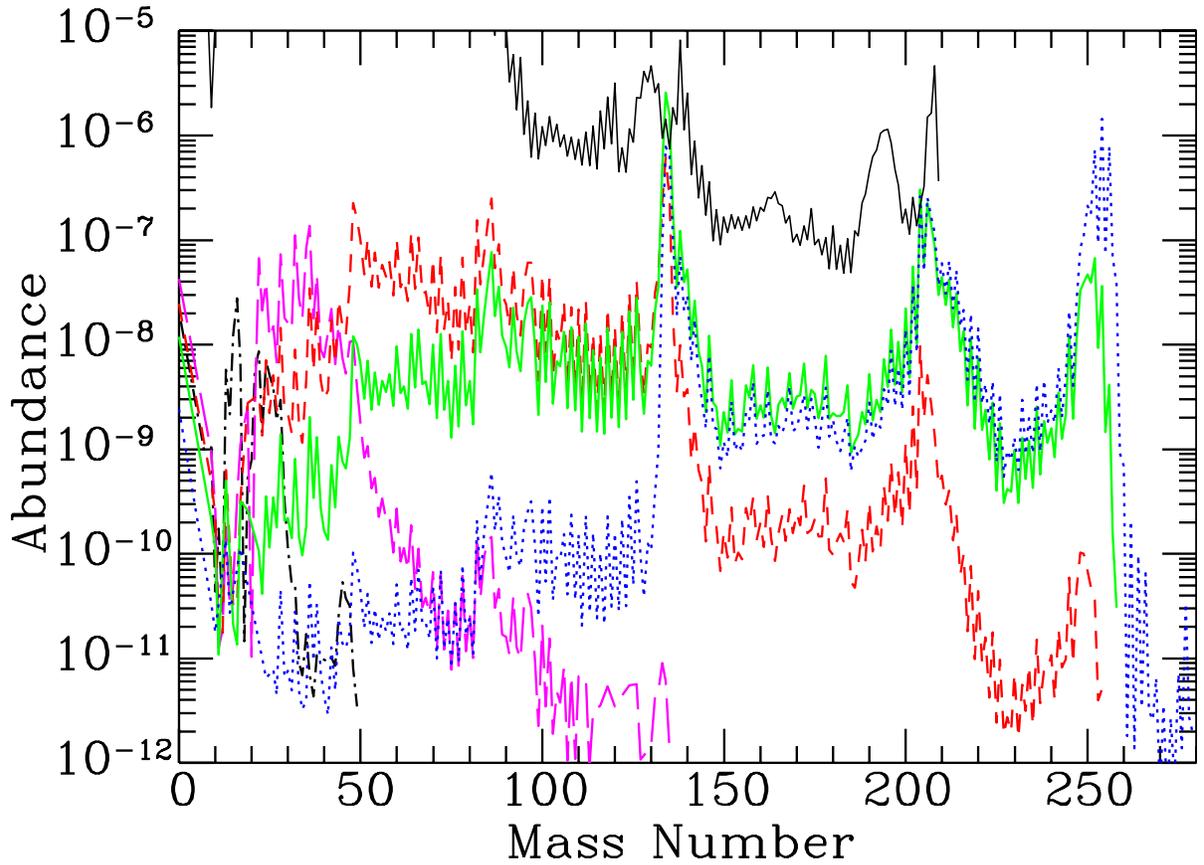} 
\caption{Final abundances by number for different $Y_e$ as a function of mass number.
The results for $Y_e =$ 0.1, 0.3, 0.4, 0.48 and 0.498 are represented
 by the dotted, solid, dashed, long-dashed and dot-dashed curves, respectively.
The solar abundance distribution in arbitrary units is shown by the uppermost, thin solid curve.}
\label{fig:heavy}
\end{figure}

Since it was found that GRB-BROs can be interesting sites of neutron capture nucleosynthesis,
 the first important comparison we can make is between the heavy element yields
 from a single GRB-BRO event and that expected from a normal core collapse SN.
For neutrino-driven wind models of SNe,
 the typical abundance by number of $r$-process peak elements in the ejecta
 is $Y_{SN\nu} \sim 10^{-4}$ at $A \ga 100$,
 and the total ejecta mass should be $M_{SN\nu} \sim 10^{-4} {\rm M_\sun}$
 (e.g. Meyer 1994, Qian 2003).
To be compared are the BRO final abundances for the realistic range of $Y_e \simeq 0.1-0.4$,
 where the peak abundance is $Y_{BRO} \sim 10^{-6}$ (Fig.\ref{fig:heavy}),
 and the total ejecta mass can be of order $M_{BRO} \sim 10^{-3}$-$10^{-2} {\rm M_\sun}$.
\footnote{
Note that for $\eta \sim 2$, $M_{BRO} \sim 10^{-2} {\rm M_\sun}$
 implies a total outflow energy $E_{BRO} \sim 4 \times 10^{52} {\rm erg}$.
This would be consistent with $L_{52}=1$
 if, for example, the GRB duration is $\Delta t \sim 10 {\rm s}$
 and the BRO is mildly collimated into a fraction $f_{\Omega} \sim 0.4$ of the sky.
Although this BRO energy is larger than
 the inferred intrinsic (i.e. beaming-corrected) total energies
 for GRB jets, $E_{GRB} \sim 10^{51} {\rm erg}$ (Frail et al. 2001),
 such values may not be so unreasonable from certain physical viewpoints
 (e.g. Levinson \& Eichler 2000),
 and is also concordant with the observed energetics of ``hypernovae'',
 which may possibly correspond to failed GRBs.
}
This gives a total heavy element yield per GRB-BRO event comparable to or somewhat less than SNe,
 albeit with large uncertainties.

As a first guess, the contribution of GRB-BROs to the solar/Galactic abundances
 then is a matter of event rates compared to SNe,
 if SNe are in fact the main sources of the solar $r$-process elements.
The intrinsic, beaming-corrected event rates
 for successful GRBs are estimated to be
 $R_{GRB} \sim 10^{-5}$-$10^{-4} {\rm yr^{-1}}$ (e.g. Frail et al. 2001, Woosley et al. 2002b),
 which is $\sim$ 0.1 - 1 \% of the SN rate $R_{SN} \sim 10^{-2} {\rm yr^{-1}}$
 (although the rates can be less for neutron star mergers; e.g. Fryer et al. 1999).
If a BRO with the above parameters is associated with every GRB as a circum-jet wind,
 then the nominal allowance of GRB-BROs to the solar abundances is $\sim$ 0.1 - 1 \%.
If failed GRBs with similar properties occur at a higher rate,
 the contribution will be even greater.
However, we must remember that the abundance patterns arising from our fiducial outflow
 are different from solar, particularly the location of the peaks.
In order to avoid overproduction of the elements between the solar $r$- and $s$-peaks,
 the true contribution may be even less.
At the very least, such GRB-BROs may be the origin of select rare elements
 not easily produced under any other conditions (Blake et al. 1981).
Alternatively, some realistic effects discussed in \S 5
 may alter the nucleosynthesis in GRB-BROs
 so as to make them potential sources of the solar $r$-process elements.

Even if GRB-BROs hold a minor share of the solar abundances,
 some observable environments may directly manifest
 the local abundance pattern of the BRO ejecta.
Detailed spectroscopic observations of metal-poor stars
 with metallicities down to [Fe/H] $\sim$ -3 have so far shown
 abundance patterns very similar to that of the solar $r$-process for elements with 56 $\le Z \la$ 70,
 at least for the so-called $r$-process enhanced stars
 (Truran et al. 2002, Qian 2003 and references therein).
This fact, together with the observed large scatter in the abundances of $r$-process elements
 as well as iron-group elements (McWilliam 1997),
 indicate that chemical evolution in the early Galaxy proceeded very inhomogeneously,
 and the abundance patterns in metal-poor stars are set mainly
 by a small number of (or even individual) SN explosions of the previous stellar generation
 (Audouze \& Silk 1995, Ryan, Norris \& Beers 1996, Shigeyama \& Tsujimoto 1998).
If GRBs are associated with the most massive range of progenitor stars,
 these explosions may have dominated in the earliest stages of the Galaxy,
 so that their nucleosynthesis products can be discerned
 in metal-poor stars with the lowest metallicities.
\footnote{
It has been argued that the observed abundance trends of iron-group elements in metal-poor stars
 at [Fe/H] $\la -2.5$ reflect explosive nucleosynthesis products of energetic ``hypernovae''
 (Nakamura et al. 2001, Umeda \& Nomoto 2002, Nomoto et al. 2002, Maeda \& Nomoto 2003).}
We can estimate the total mass of gas swept up by a GRB-BRO
 and with which the BRO ejecta mixes to be $M_{sh} \sim 3 \times (10^4$-$10^5) {\rm M_\sun}$
 (Ryan et al. 1996, Shigeyama \& Tsujimoto 1998).
The dilution factor per event is then $f_{MPS} = M_{BRO}/M_{sh} \sim 10^{-7.5}$,
 and the expected heavy element abundance is $Y_{MPS} \sim f_{MPS} Y_{BRO} \sim 10^{-13.5}$
 or roughly 2.5 dex below solar, an observable level.
It is intriguing to see if future observations of extremely metal-poor stars
 (e.g. Christlieb et al. 2002)
 reveal a heavy element abundance pattern distinct from that of the solar $r$-process.

Another interesting locale is the companion star in a black hole binary system,
 where the explosive event that formed the black hole
 may have left traces of the explosion ejecta on the surface of the companion.
(This argument applies to massive star core collapse models like collapsars and supranovae,
 but excludes neutron star merger scenarios.)
Indeed, observations of the companion stars of GRO J1655-40 (Nova Sco; Israelian et al. 1999)
 and other black hole binaries (Orosz et al. 2001, Israelian et al., in preparation)
 have uncovered anomalous $\alpha$-element enhancements
 indicating contamination by supernova ejecta, possibly of the energetic ``hypernova'' variety
 (Israelian et al. 1999, Podsiadlowski et al. 2002).
Very crudely, the fraction of the BRO ejecta captured by the companion may be
 $f_{cap} \sim 10^{-3}$, considering a typical solid angle subtended by the star,
 provided that it lies within the (presumably wide) collimation cone of the BRO.
The mass of the mixing zone between the ejecta and the companion surface
 could be in the range $M_{mix} \sim 10^{-4}-10^{-2} {\rm M_\sun}$ (Israelian et al. 1999, Qian 2000),
 so that the dilution factor here is $f_{bin} \sim f_{cap} M_{BRO}/M_{mix} \sim 10^{-4}-10^{-1}$,
 although these values are very uncertain.
This leads to $Y_{bin} \sim f_{bin} Y_{BRO} \sim 10^{-10}-10^{-7}$,
 which is a staggeringly high overabundance with respect to solar $Y_{\sun} \sim 10^{-11}$,
 and should be clearly detectable.
High resolution spectroscopic observations of such stars are warranted 
 to search for neutron-capture elements and study their detailed abundance patterns.

In discerning light element products from BROs,
 the only hope is offered by the black hole binary companions,
 as their contribution to the solar abundances
 and even very metal-poor stars can be shown to be miniscule.
Taking the $\eta=2$ case, we will be most optimistic
 and assume the least dilution $f_{bin} \sim 10^{-1}$,
 and also select $Y_e$ where the maximum final abundance is achieved for each element.
For D, $^7$Li, $^6$Li, $^9$Be and $^{11}$B, respectively,
 these are $Y_e \sim$ 0.1, 0.7, 0.5, 0.4 and 0.1,
 at which the abundance by number on the companion surface would be
 $Y_{bin} \sim 10^{-3}, 2 \times 10^{-8}, 10^{-10}, 10^{-10}$ and $4 \times 10^{-8}$.
This must be compared to the solar abundances for the respective elements,
 $Y_{\sun} \sim 4 \times 10^{-5}, 10^{-8}, 6 \times 10^{-10}, 10^{-10}$ and $10^{-9}$.
Overabundance is expected to be high only for D, and is marginal for $^7$Li and $^{11}$B.
More dilution or deviations from the optimal $Y_e$,
 as well as stellar depletion effects that can be severe for D,
 would render them difficult to observe, if at all.
However, light element production may become more prolific
 for BROs which are nonrelativistic and/or collimated,
 or when circumstances allow for significant spallation (\S 5).

Either in extremely metal-poor stars or in black hole binary companions,
 observations of the pure ejecta abundance pattern 
 may provide us with valuable insight into the origin of GRBs.
It is evident in Figs.\ref{fig:light} and \ref{fig:heavy}
 that the final abundance patterns of both the light and heavy elements
 are quite sensitive to $Y_e$,
 which in turn implies that observations can significantly constrain its value.
Since $Y_e$ is an important quantity
 characterizing the nature and conditions of the central engine,
 the nucleosynthetic products can shed light
 on the innermost regions of GRB sources, which are otherwise very difficult to probe.

\section{Conclusions and Further Prospects}

The salient points of this work are summarized.
We have explored the possibility of nucleosynthesis
 in baryon-rich outflows that are expected to accompany GRBs.
Detailed reaction network calculations were performed
 in the context of the fireball model with varying baryon load.
It was found that mildly relativistic outflows
 can synthesize appreciable quantities
 of heavy neutron capture elements with masses up to the actinides via the ``$n$-process'',
 as well as interesting amounts of light elements such as deuterium.
Their characteristic abundance patterns may be discernible in
 black hole binary companion stars or in extremely metal-poor stars,
 possibly offering important information on the GRB central engine conditions.
Some contribution to the solar heavy element abundances may also exist.

Our intention here was to carry out a pilot study,
 focusing on the most general and model-independent aspects
 of nucleosynthesis in BROs related with GRBs.
In reality, various different types of GRB-BROs can arise (\S 1),
 and the present calculations may be fairly indicative for some kinds of BROs, but not for others.
For accretion disk winds that originate in regions and by mechanisms distinct from the GRB jet,
 the fireball model may not be so suitable for describing their basic properties
 such as the temperature, entropy and expansion timescale (\S 2).
Likewise, the characteristics of BROs occurring in failed GRBs
 are dependent on the particular manner in which GRB jet fails.
Detailed studies of nucleosynthesis in realistic GRB-BRO conditions
 require a good understanding of both the dynamics and microphysics of the true GRB engine.
Since we are still far from this,
 one approach for undertaking more in-depth investigations
 is to work in the framework of specific models for the GRB progenitor and outflow formation.
Alternatively,
 we can also make progress by considering in steps various physical aspects
 that can be generically important for a range of possible GRB scenarios.
Below we give brief discussions of some effects
 that have not been accounted for in this work,
 but which may have significant consequences for the outcome of GRB-BRO nucleosynthesis.
These considerations point us toward a variety of subsequent studies.

In our reaction network calculations,
 fission was not included for the sake of simplicity.
However, we saw that for a plausible range of $Y_e$
 the reaction flow can extend well into the actinide region
 where fission should be unavoidable.
Since the free neutrons in our BROs always stay abundant until very late times (\S 3.2),
 a highly probable consequence is fission cycling,
 whereby fission products act as seed nuclei
 for successive cycles of reaction flows into the fission regime.
This should affect the resulting heavy element abundances in interesting ways
 and is currently under investigation (Terasawa et al., in preparation).
Another facet not included in our reaction network
 is heavy nuclei in the proton-rich region,
 which should be important for situations when $Y_e \ga 0.5$
 and may potentially give rise to $p$-process nuclei
 (Iwamoto et al., in preparation; see also Meyer 2002).

Our dynamical framework here was
 limited to the basic spherical fireball model.
Although the BROs may not be as strongly collimated as the GRB jets,
 some degree of collimation
 (e.g. through external gas pressure gradients or magnetic stresses)
 is naturally expected, which modify in the density and temperature trajectories
 from eqs.\ref{eqn:trajacc} and \ref{eqn:trajcoa}.
While the details are model-dependent,
 collimation should generally tend to prolong the phase of initial expansion
 when the charged particle reactions mainly proceed
 (Lemoine 2002, Beloborodov 2003b).
This would permit the initial reaction flow to reach larger $A$,
 resulting in increased heavy element production for cases with high neutron excess,
 as well as larger light element yields.
Magnetic fields can also crucially modify the flow acceleration in addition to collimation,
 but assessing their effect on nucleosynthesis requires detailed dynamical modeling.

For reasons discussed in \S 2,
 we had also restricted the dimensionless entropy to $\eta \ga 2$,
 meaning that the outflows are at least mildly relativistic
 and the total baryon masses (isotropic equivalent) are $M_b \la 2.8 \times 10^{-2} {\rm M_\sun} E_{53}$.
However,
 much larger baryon loading can conceivably happen in GRB models that activate outflows inside massive stars
 (MacFadyen \& Woosley 1999, MacFadyen 2003, Wheeler et al. 2002),
 leading to BROs in the nonrelativistic regime.
These should be even more efficient nucleosynthetic cauldrons,
 as they involve a lower entropy,
 longer initial expansion time, and larger total ejecta mass.
In fact, the effects of nonrelativistic baryon loading (and to a lesser extent, collimation)
 may allow the realization of physical conditions
 that are ideal for reproducing the solar $r$-process abundances (Pruet et al. 2003).
This interesting prospect of GRB-BROs as potential $r$-process sites
 is worth exploring in some detail.
(See below for a discussion of previous work on the $r$-process in magneto-rotational jet-powered supernovae.)

Another significant but model-dependent process
 is neutrino irradiation of the BRO from a central source.
Strong thermal neutrino emission is expected in most types of GRB progenitors,
 including massive star core collapse
 and neutron star mergers.
This can impact the nucleosynthesis in different ways,
 either through its thermal and dynamical effects on the BRO expansion,
 or through neutrino-induced interactions on the nucleons and nuclei
 (Meyer 1995, Meyer, McLaughlin \& Fuller 1998).
Relatively realistic calculations including the effects mentioned so far
 may be a feasible goal for nucleosynthesis in the baryonic winds of neutron star mergers,
 employing outflow trajectories of density and temperature calculated in numerical simulations
 (Janka \& Ruffert 2002).

The considerable fraction of free neutrons remaining after BRO nucleosynthesis (\S 3.2)
 may induce some noteworthy effects
 before they decay into protons at a radius $r_{n} \sim 3 \times 10^{13} {\rm cm} \Gamma$.
For nucleosynthesis,
 an intriguing possibility is the operation of an ``external'' neutron capture process,
 i.e. heavy element synthesis by capture of the BRO neutrons
 on any external baryonic matter with which the outflow interacts and mixes,
 especially when the BRO is nonrelativistic.
The external material may be the outlying stellar wind of the progenitor star,
 the surface of a binary companion,
 or the overlying stellar envelope in the case of a collapsar (Pruet et al. 2003).
Such processes may be externally seeded,
 as any nuclei like Fe preexisting in the environs
 can constitute the seed for neutron captures.
Further interesting consequences of neutrons can be caused by
 their kinematic decoupling from the rest of the fireball plasma,
 such as modifications to the dynamical evolution of the fireball and/or blastwave
 (e.g. Derishev et al. 1999, Fuller et al. 2000, Beloborodov 2003a),
 characteristic neutrino and gamma-ray signatures
 (Bahcall \& M\'esz\'aros 2000, M\'esz\'aros \& Rees 2000),
 and production of light elements by spallation (Pruet et al. 2002; see below).

Although we only dealt with thermal nuclear reactions in the expanding BRO,
 additional, nonthermal nuclear processes can alter the nuclear composition in different ways in real GRB-BROs.
These may actually dominate the production of some elements,
 particularly D, Li, Be and B.
For the high $\Gamma$ GRB jets,
 the observed short timescale variability of the GRB emission
 may entail substantial fluctuations and dissipation by internal shocks within the outflow,
 which can lead to spallation of $^4$He and other existing nuclei (Beloborodov 2003b).
\footnote{
This effect is probably less important for the low $\Gamma$ BROs studied here,
 since they may not be as strongly time-varying as GRB jets,
 and also because internal shocks in the BROs would occur at much smaller radii
 where the temperature is still high and the outflow optically thick.
}
Appreciable spallation may also be induced
 by interaction of the outflow with external matter
 (e.g. progenitor stellar wind, binary companion star, massive star envelope),
 affecting both the nuclei inside the BRO and those composing the external baryons (Cameron 2001).
Under circumstances where free neutrons become kinematically decoupled,
 bulk motions relative to the rest of the flow may develop
 and induce spallation by neutrons, either internally or externally 
 (Pruet et al. 2002, Beloborodov 2003b).
\footnote{
In our BROs, however, coupling of the neutrons to the rest of the plasma
 through elastic n-p collisions generally persist until late times.
}
Photodisintegration by nonthermal gamma-rays emitted in internal shocks
 can also take place outside the photosphere (Lemoine 2002),
 although this may not be so significant for BROs due to their large photospheric radii.

Virtually all of the above effects (plus some other unique ones)
 may become essential for scenarios
 in which the GRB jet and associated BRO
 initially propagate through the dense envelope of a massive star
 before emerging and generating a GRB (Woosley 1993, Woosley et al. 2002b, Wheeler et al. 2002).
There is a multitude of deviations from the simple fireball picture
 due to the strong outflow-star interaction,
 such as reverse shock heating of the flow,
 expansion of the shocked material into a cocoon,
 entrainment and mixing with the stellar material through instabilities,
 breakout from the envelope and ensuing lateral expansion, etc.
 (MacFadyen \& Woosley 1999, MacFadyen et al. 2001,
 M\'esz\'aros \& Rees 2001, Ramirez-Ruiz et al. 2002).
In fact, the expansion timescale at envelope breakout (radii $r \sim 10^{11} {\rm cm}$)
 may be more important for nucleosynthesis than the expansion timescale at the central engine
 (see discussions in Pruet et al. 2002, 2003).
These effects may also be highly variable and inhomogeneous in nontrivial ways
 (Aloy et al. 2000, 2002, Zhang et al. 2003).
Quantitative modeling of nucleosynthesis within this complicated picture
 is a very interesting but challenging goal.
Note that the ``disk winds'' in collapsars have been suggested
 to be an important site for synthesis of $^{56}$Ni,
 which is essential for powering the light curve of associated supernovae/hypernovae
 (MacFadyen \& Woosley 1999, Pruet et al. 2003, MacFadyen 2003).

The physical properties of the BROs considered here
 may actually bear a number of similarities to the magneto-rotational ``jets'' in core collapse supernovae
 originally studied by LeBlanc \& Wilson (1970).
If the collapsing core has sufficiently fast rotation and strong magnetic fields,
 such jets (or related mechanisms) may potentially power the supernova explosion
 (Bisnovatyi-Kogan 1971, Meier et al. 1976, Symbalisty 1984, Khokhlov et al. 1999, Wheeler et al. 2002).
LeBlanc-Wilson jets have also been discussed as promising $r$-process sites
 (Schramm \& Barkat 1972, Symbalisty, Schramm \& Wilson 1985, Cowan, Thielemann \& Truran 1991),
 although no detailed nucleosynthesis calculations in this picture have been conducted.
They are expected to expel $\la 10^{-2} {\rm M_\sun}$ of highly neutron-rich material
 from within the proto-neutron star as narrowly collimated ejections (Symbalisty 1984).
In the context of our fireball model,
 this should correspond roughly to an isotropic equivalent baryon load of $\la 1 {\rm M_\sun}$.
A significant fraction of the jet energy is also in magnetic form.
In comparison, our fiducial $\eta=2$ outflow
 with isotropic equivalent mass $\simeq 10^{-2} {\rm M_\sun} E_{53}$ and without magnetic fields
 has much lower density, higher entropy and shorter expansion timescale,
 leading to an $n$-process rather than an $r$-process.
In light of the growing observational evidence for asymmetric supernova ejecta,
 ultra-strong magnetic field neutron stars and potential connection with GRBs
 (Wheeler et al. 2002 and references therein),
 it should be worthwhile to reconsider in greater depth
 the physics and implications of such magneto-rotational outflows in core collapse supernovae
 (Woosley et al. 2002a,b, Wheeler et al. 2002).
Note that a related $r$-process scenario has been discussed recently by Cameron (2001),
 where an $r$-process involving fission cycling takes place inside an accretion disk,
 and the products are subsequently ejected in jets.

The ultimate answer to the problem of GRB-BRO nucleosynthesis is contingent
 on elucidating the physics of the GRB central engine, including
 the progenitor's identity, the central engine configuration
 and the outflow formation mechanism, from which we may be a long way.
Nevertheless, we may hope to build increasingly realistic models,
 adding step by step different levels of complexity.
Comparison of such calculations to observations of various sites
 may then allow us to constrain the unknown properties of the GRB source,
 as well as to clarify the role of GRB-BROs in the chemical evolution of the Galaxy.
Further investigations of GRB-BRO nucleosynthesis in various directions
 should provide us with a rich scope of interesting work for the future.

\acknowledgments
We are very grateful to the referee, Stan Woosley,
 for helpful and stimulating discussions before and after submission of this paper.
Many thanks also go to Thomas Janka for numerous insightful conversations,
 as well as Grant Mathews, Brad Meyer, Wolfgang Hillebrandt, Claudia Travaglio and Hideyuki Umeda
 for helpful comments.
Takahiro Sasaqui is acknowledged for his help in the early stages of this study.
One of the authors (M. T.) wishes to acknowledge the fellowship
 of the Japan Society for Promotion of Science (JSPS).

\clearpage

\end{document}